# Observation of out-of-plane vibrations in few-layer graphene

Chun Hung Lui[1], Leandro M. Malard[1], SukHyun Kim[1], Gabriel Lantz[1],
François E. Laverge[1], Riichiro Saito[2], Tony F. Heinz[1*]

*[1]Departments of Physics and Electrical Engineering, Columbia University, 538 West 120th Street, New York, NY 10027, USA*
*[2]Department of Physics, Tohoku University, Sendai 980-8578, Japan*

We report the observation of layer breathing mode (LBM) vibrations in few-layer graphene (FLG) samples of thickness from 2 to 6 layers, exhibiting both Bernal (AB) and rhombohedral (ABC) stacking order. The LBM vibrations are identified using a Raman combination band lying around 1720 $cm^{-1}$. From double resonance theory, we identify the feature as the LOZO' combination mode of the out-of-plane LBM (ZO') and the in-plane longitudinal optical mode (LO). The LOZO' Raman band is found to exhibit multiple peaks, with a unique line shape for each layer thickness and stacking order. These complex line shapes of the LOZO'-mode arise both from the material-dependent selection of different phonons in the double-resonance Raman process and from the detailed structure of the different branches of LBM in FLG.

PACS numbers: 78.30.-j; 63.20.D-; 71.38.-k; 61.46.-w

* Corresponding author: tony.heinz@columbia.edu

The low-energy electronic properties of few-layer graphene (FLG) and, accordingly, the transport and infrared optical properties are defined by the interactions between the graphene layers. This topic has been focus of much recent research. The influence of the interlayer interactions on the *vibrational properties* of FLG is also a topic of great interest. The different layer-breathing modes (LBMs), which involve out-of-plane relative displacement of individual graphene layers, are of particular importance because of their direct sensitivity to layer number and stacking order. Though LBMs have been investigated in the bulk graphite [1, 2] and multi-walled carbon nanotubes [3, 4], studied theoretically in FLG [5-13], and reported in a recent experiment of trilayer graphene (3LG) [14], it has yet to be comprehensively investigated in FLG samples with different thickness and stacking sequences.

In this paper, we report the observation of the LBM in graphene samples with thickness from 2 to 6 layers with both Bernal (AB) and rhombohedral (ABC) stacking. We observe this mode, which we designate as the ZO' mode, through its signature in the LOZO' combination Raman band formed with the in-plane longitudinal optical (LO) phonon. The measured LOZO' frequency is found to shift with the photon energy ($E_{exc}$) of the excitation laser, a consequence of the selection of different phonon wave vectors in the electronically resonant two-phonon Raman process [15-17]. Comparison of the Raman dispersion with that of the two underlying phonons in the combination band provides a clear signature of the physical origin of the observed response. A striking feature of the LOZO'-mode is its marked sensitivity to both the layer thickness and stacking order of the FLG material. With increasing layer thickness the LOZO' spectra exhibit additional peaks, which shift systematically in frequency. Analysis of the spectra reveals the role played by both the vibrational and electronic properties of FLG. The frequencies of the LBM phonons are, of course, reflected in those of the combination modes, but the electronic structure is also important, since it determines the in-plane wave vectors of the phonons through the doubly resonant Raman process. The measured spectra thus contain fundamental information about the electron and phonon band structure of FLG, as well as about the electron-phonon (*e-ph*) interactions, which determine the strength of the features. From the standpoint of material characterization, the LOZO' Raman mode provides a simple and unambiguous optical method for the identification of both the layer number and stacking order for FLG samples of up to 6 layers in thickness. The approach thus complements and extends use of the 2D Raman mode [14, 16-18].

In our experiment, we investigated pristine mechanically exfoliated graphene layers that were freely suspended over trenches, as well as samples deposited directly on fused quartz or $SiO_2/Si$ substrates [19]. We identified the layer thickness and stacking order by means of infrared spectroscopy [18, 20-23] for graphene samples with layer number $N$ = 1 to 6 with both AB and ABC stacking order [23]. The Raman measurements were carried out using a commercial (JY Horiba) confocal microRaman system under ambient condition [19] over the spectral range of 1625 to 2100 $cm^{-1}$. We note that there were no systematic differences in the mode frequencies for the suspended and supported samples, although weak features were typically somewhat sharper and easier to identify for the suspended samples.

Fig. 1 shows the Raman spectrum for free-standing monolayer and bilayer graphene (1LG, 2LG), as well as for graphite sample [14, 24-26]. The Raman intensity in this range is roughly two orders of magnitude smaller than that of the G mode. We observe a weak (and as-yet unreported) Raman feature at ~1655 $cm^{-1}$ and two stronger peaks at ~1730 and ~1765 $cm^{-1}$ [24-26]. We denote these peaks, respectively, as P1, P2 and P3. None of these three Raman peaks is observed for suspended 1LG or for 1LG supported on quartz substrates [25, 26] for different $E_{exc}$ [19]. As shown in Fig. 2 for 2LG, the peaks are found to exhibit dispersion. While the frequency of P3 changes only slightly with $E_{exc}$, both P1 and P2 significantly blue shift with $E_{exc}$, with average energy dependence of 38 and 29 $cm^{-1}$/eV, respectively. As we describe below, we have applied double-resonance Raman theory [15-17, 27] to analyze the dispersion behavior and to assign P1, P2, and P3, respectively, to the LO+ZA, LO+ZO' combination modes, and to the 2ZO overtone mode by an intra-valley resonant scattering process. Here ZA and ZO denote, respectively, the out-of-plane acoustic and optical modes of a graphene layer near Γ point in the Brillouin zone. With the above assignments, the absence of P1-P3 in 1LG can be readily understood: the ZA and ZO mode are not Raman active in SLG [8], while the LBM (ZO') obviously requires more than one graphene layer. We focus on the LOZO' feature and ZO' mode in this paper; the LOZA and 2ZO modes are discussed in the supplemental material [19].

In our analysis of P2 (LOZO') in 2LG, we investigated the phonon dispersion of the LO and ZO' branch experimentally by measuring directly their overtone modes (2LO and 2ZO') [19]. Both overtone modes exhibit dependence on $E_{exc}$ arising from the double-resonance Raman scattering mechanism. In particular, we observed two components ($2ZO'^{+}$ and $2ZO'^{-}$) in the 2ZO' line [19], which we attribute to resonances between different electronic bands in 2LG. Fig. 2(b) shows the LO and ZO' phonon energies, obtained as half of their overtone values. We found that the combination of a LO phonon and a ZO' phonon matches well with the frequency and dispersion of feature P2. Fig. 3(a) further displays the ZO' phonon energies obtained by subtracting the LO phonon energies from the LOZO'-mode energies as a function of phonon momentum. The results are compared with theoretical predictions [12] and with the ZO' phonon energies obtained from the 2ZO' overtone mode. The good agreement with both the experimentally and theoretically derived values confirms our assignment of P2 as the LOZO' combination mode. (Figs. 2 and 3(a) also display the results of similar analysis on LOZA and 2ZO mode, with detailed discussion provided in the supplemental material [19].) Our assignment is also consistent, both in terms of the Raman shift and dispersion characteristics, with a recent theoretical calculation of the LOZO' spectra using the double-resonance Raman theory [13].

Since the LOZO' mode involves the layer-breathing vibrations, its behavior should be sensitive to interlayer interactions. To explore this issue, we measured the LOZO' line for graphene samples of layer thickness $N$ = 2 to 6 (2LG to 6LG), with both AB and ABC stacking sequence [Fig. 4(a)]. The LOZO' band exhibits multi-peak features in all samples. The positions of the sub-peaks of the LOZO' band evolve systematically, increasing from 2LG to 6LG. Representing the frequency by center of mass of the LOZO' band, we find that the average frequency for FLG with AB (ABC) stacking increases by ~18 $cm^{-1}$ (22 $cm^{-1}$) from 2LG to 6LG and approaches (slightly exceeds) the value of the bulk Bernal graphite. As the in-plane LO mode is essentially unchanged by interlayer coupling [6], the observed layer-dependent behavior of the LOZO' mode is expected to reflect the behavior of the LBM (ZO'), which is known to be sensitive to the layer thickness.

In addition to the layer-dependent behavior, the LOZO' mode is also exhibits strong sensitivity to the stacking sequence (relevant for $N$ = 3 to 6). In the AB spectra [Fig. 4(a)], the peak at the highest frequency is significantly stronger than the other sub-peaks. As $N$ increases, the number of sub-peaks increases and their positions shift towards the main peak. In contrast, the ABC spectra [Fig. 4(a)] generally exhibit more peaks, with narrower line widths and more evenly distributed spectral weight, than the AB spectra. The energies of peaks in the ABC spectra are also differ somewhat from those in the AB spectra. Such distinctions between the AB and ABC spectra are found to persist for all $E_{exc}$ (see for example the case of 3LG in the supplementary materials [19]).

The sensitivity of the Raman LOZO' spectrum to layer number and stacking order arises from two factors. First, the distinct electronic structure of different FLG samples influences the Raman response through the selection of different phonon wave vectors by the double resonance process. This mechanism is responsible for the layer and stacking-order dependence of the line shape of the 2D Raman mode [14, 16-18]. Second, the layer thickness and stacking order of the FLG directly affect the ZO' modes in the material and, hence, alter the characteristics of the LOZO' band. In particular, for $N$-layer graphene, there are $N$-$1$ distinct ZO' modes (which we label as ZO', ZO'', etc.), each with a different frequency. For the 2D Raman mode, the first mechanism is critical for its sensitivity to layer and stacking-order, but the second is insignificant since the relevant in-plane phonons in 2D mode are hardly altered by the interlayer interactions.

To clarify the roles of these two mechanisms for the LOZO' mode, let us first analyze 2LG. Already in this case, the LOZO'-mode exhibits a somewhat complex line shape, with a pair of peaks at 1716 $cm^{-1}$ and 1727 $cm^{-1}$ [Fig. 4(a)]. These two peaks do not arise from the phonon band structure: 2LG has only one LBM and the splitting of the two branches of the LO mode is small (~2 $cm^{-1}$) [6]. We therefore examine different possible electronic processes for resonant Raman scattering to explain the behavior. For this purpose, we treat the 2LG electronic structure within a simple tight-binding model, including only the dominant intralayer ($\gamma_0$) and interlayer ($\gamma_1$) couplings. We can then determine the minimum ($k_{min}$) and maximum ($k_{max}$) phonon momenta at $E_{exc}$ = 2.33 eV [Fig. 3(a)] for resonant processes corresponding to the intra-band scattering within the low- and high-lying conduction (valence) bands. Using these two momenta and the theoretical phonon dispersion relations shown in Fig. 3(a), we find the sum of the energies of the corresponding LO and ZO' phonons differ from one another by ~15 $cm^{-1}$, which is comparable to our experimentally observed separation of 11 $cm^{-1}$ between the two LOZO' components. We therefore ascribe the two peaks in the LOZO' mode of 2LG to the different electronic resonances in the two-phonon Raman scattering processes.

The same mechanism appears to be responsible for some of the structure in the LOZO' band in ABC stacked 3LG [Fig. 4(a)]. For this case, the spectrum exhibits three peaks, separated from one another by ~20 $cm^{-1}$. On the other hand, there are only two LBMs, with a separation of ~45 $cm^{-1}$ [Fig. 3(b)]. With respect to electronic resonances, we follow the same analysis as for 2LG to define phonon momenta $k_{max}$ and $k_{min}$ [Fig. 3(b)] for different electronic resonances for the electronic structure of ABC-3LG. We find that the sum of energies of the corresponding LO and ZO' states for $k_{max}$ and $k_{min}$ differ by 20 $cm^{-1}$, consistent with the experimental observation for the peak splitting.

Our experimental data also provide evidence for the influence of the various LBM branches in FLG in defining the peaks in the LOZO' band. In particular, for FLG with $N > 3$, the separation between the peaks within the LOZO' band become comparable to the energy splitting of the LBM branches. Also the frequencies of all the sub-peaks, as shown in Fig. 4(b), evolve systematically with the layer number in a way that is analogous to the behavior of the LBM branches in FLG. For a more quantitative analysis, we estimate the energies of the LBM branches in FLG within a scheme of graphene layers interacting with only (identical) coupling between adjacent graphene layers. Since adjacent graphene layers have identical stacking for both AB and ABC stacking orders, we predict the same behavior for both stacking orders, namely, $\omega_n(N) = \omega_o \sin(n\pi/2N)$, where the integer $0 < n < N$ specifies the branch of the mode, $\omega_o$ is the frequency of the LBM mode for the bulk graphite, and the frequencies correspond to those for zone-center phonons. Here we set $\omega_o$ = 132.3 $cm^{-1}$, taken as half of the measured value of the double-resonant 2ZO' overtone mode for graphite. As dispersion of the LBM is rather weak near the zone center (Fig. 3), the inaccuracy of our estimation from the in-plane phonon dispersion is small (~2 $cm^{-1}$). The calculated $\omega_n(N)$, upshifted to account for the LO phonon, are displayed in Fig. 4(b). The results reproduce the trends and approximate frequencies of the different LOZO' sub-peaks for the FLG samples. Some deviations are observed for the high-energy peaks of the ABC spectra, which scatter on the upper and lower sides of the theoretical curve. We assume that they are associated with the same LBM branch, but different electronic resonances and we take the average of these values to compare with the simple model. Although more rigorous calculations are needed to investigate the detailed mechanism of the LOZO' mode, our analysis strongly suggests that different LBM branches present for layer thickness $N > 2$ can contribute to the observed LOZO' bands in FLG.

In conclusion, we have identified the LOZO' combination mode Raman band for FLG, thus gaining access to the low-frequency layer breathing (ZO') mode. The Raman band is found to be sensitive to interlayer interactions and exhibits a unique line shape for graphene of each layer thickness and stacking order. The distinctive properties of LOZO' mode reflect the roles of both the different LBM branches and the selection of phonon wave vectors in the double resonance Raman process. The LOZO' mode is a convenient probe of the low-energy out-of-plane vibrations and provides an effective tool to characterize the layer number and stacking order in FLG.

We thank Y. Miyauchi and K. Sato for discussions and Z. Q. Li for assistance in sample preparation. RS acknowledges MEXT grant (No. 20241023). LMM acknowledges a CNPs Fellowship. Research at Columbia University was supported by the Office of Naval Research through the MURI program.

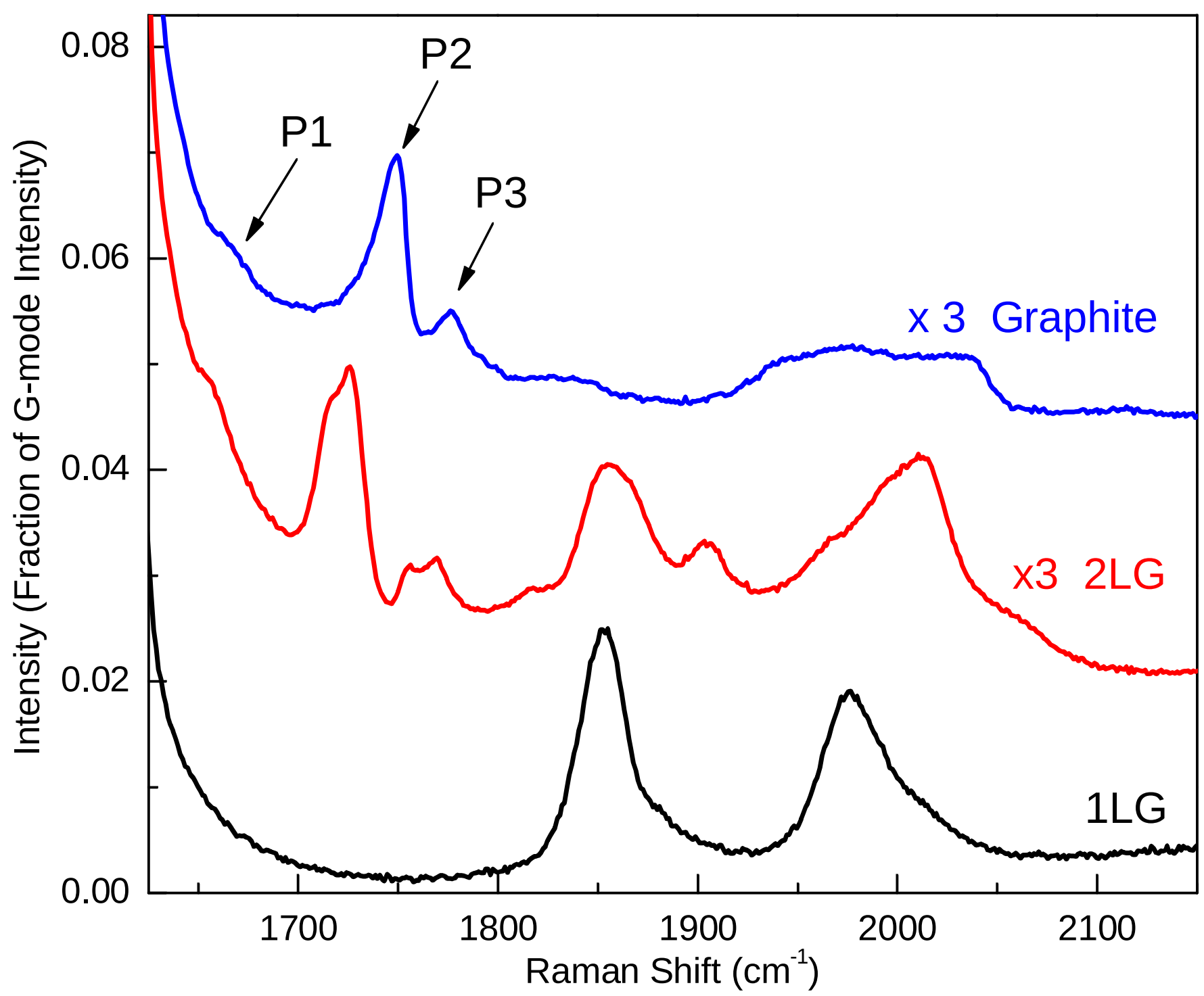

FIG. 1. Raman modes in the range of 1625 to 2150 $cm^{-1}$ for free-standing 1LG and 2LG, compared with bulk graphite for $E_{exc}$ = 2.33 eV. The spectra are normalized with respect to the G-mode peak intensity. The spectra of 2LG and graphite are increased by a factor of 3 in magnitude and displaced by 0.02 for clarity.

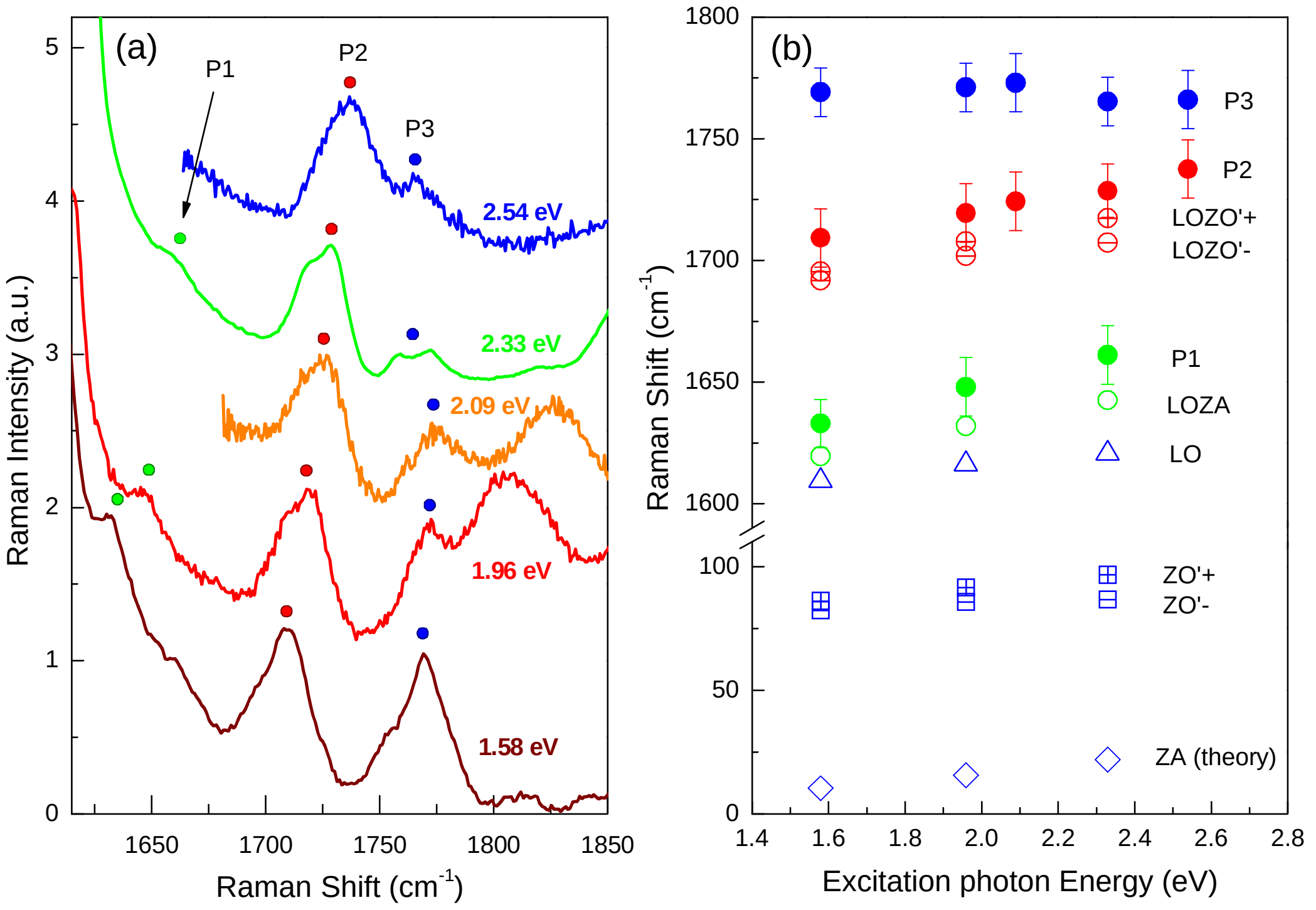

FIG. 2. (a) Raman spectra of 2LG (displaced for clarity) in the range of 1615 to 1850 $cm^{-1}$ for different $E_{exc}$. For $E_{exc}$ = 1.96 and 2.33 eV, the measurements were performed on free-standing 2LG; for the other cases, the other samples were supported on fused quartz substrates. Note that the strongly dispersing LO+TA band (at ~1810 $cm^{-1}$ for $E_{exc}$ = 1.96 eV) overlaps with P3 for $E_{exc}$ =1.58 eV. (b) The frequency (solid dots) of different Raman peaks as designated in (a) as a function of $E_{exc}$. The circles are comparisons based on the appropriate combination of the different component phonon modes in the LOZO' mode. The LO and ZO' phonon frequencies were determined experimentally from their overtone modes [19]. The diamonds are theoretical values of ZA phonon frequency from Fig. 3(a).

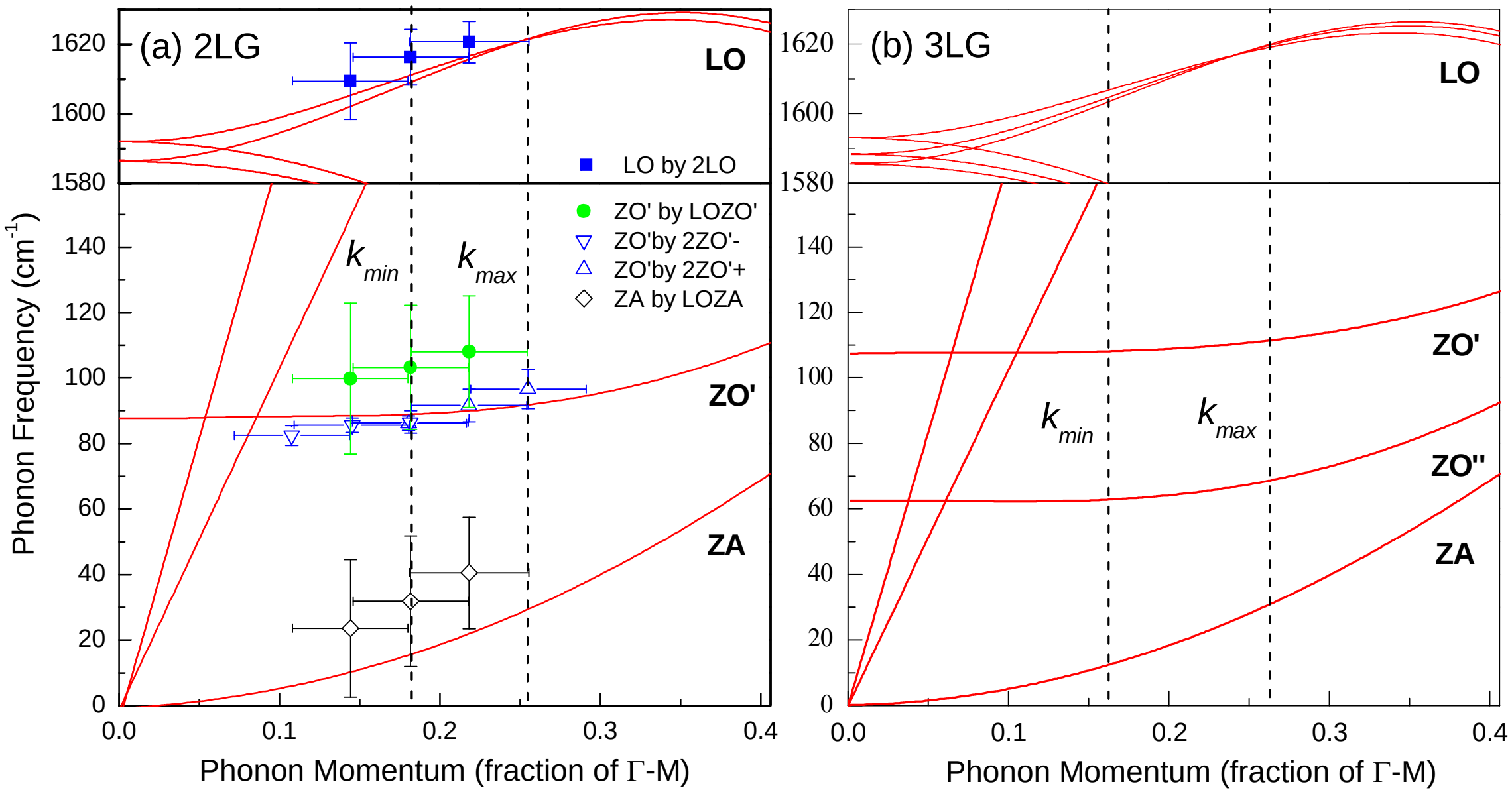

FIG. 3. (a) Comparison of experimental (symbols) and theoretical (lines) frequencies for the LO, ZO' and ZA modes in 2LG. The LO (squares) and ZO' (triangles) phonon frequencies were determined experimentally from their overtone modes [19]. The ZO' (dots) and ZA (diamonds) phonon frequencies were obtained by subtracting the LO phonon frequencies (squares) from the measured LOZO'- and LOZA-mode frequencies, respectively. The errors bars represent uncertainties defined by the width of the Raman lines. The phonon momenta for the experimental points are taken as the average transition momenta in 2LG, with the error bars representing the uncertainties arising from various resonant Raman scattering processes. (b) Same as (a) for 3LG. We here neglect the next-nearest-layer interaction and assume the phonon band structure to be the same for ABA and ABC trilayers. In both (a) and (b), the theoretical curves for LO modes are from [6], while those for low-energy modes are from [12]. The dashed lines are representative phonon momenta in the double-resonance processes for 2LG and 3LG at $E_{exc}$ = 2.33 eV, as described in the text.

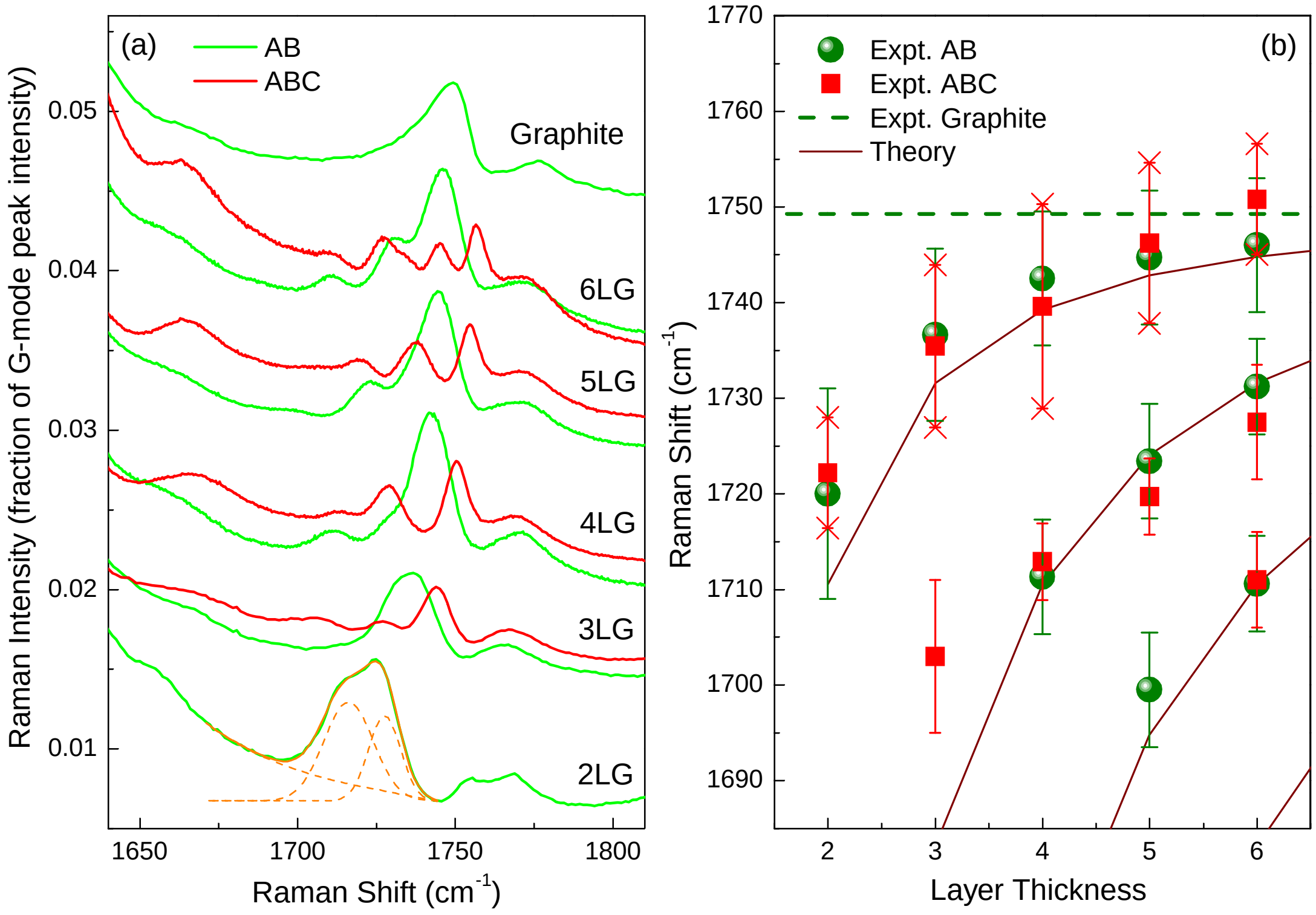

FIG. 4. (a) Raman spectra in the range of 1640 to 1810 $cm^{-1}$ for FLG of different thickness and stacking order with $E_{exc}$ = 2.33 eV. We used free-standing 2LG, when available. For thicker FLG, the samples were supported on $SiO_2$/Si substrates. The spectra are normalized with respect to the G-mode peak intensity. The orange lines represent a double-Gaussian fit for the 2LG LOZO' peak. (b) The energy of the sub-peaks in the LOZO' band for FLG with two stacking orders as a function of layer thickness. The errors bars correspond to the width of the Raman features. For the high-energy peaks of the ABC spectra, we present the average values of the two highest energy peaks. The positions of the original peaks lie at the ends of the corresponding error bars. For 2LG, the green dot is the average energy, and the red crosses and squares are, respectively, the peak positions of the two Gaussian components and their average value. The lines are the calculated energies of the LBM phonons described in the text, but up shifted to match the value in the limit of bulk graphite (dashed line).

# Supplemental Material of

# "Observation of out-of-plane vibrations in few-layer graphene"

Chun Hung Lui[1], Leandro M. Malard[1], SukHyun Kim[1], Gabriel Lantz[1], François E. Laverge[1], Riichiro Saito[2], Tony F. Heinz[1]

[1]*Departments of Physics and Electrical Engineering, Columbia University, 538 West 120th Street, New York, NY 10027, USA*

[2]*Department of Physics, Tohoku University, Sendai 980-8578, Japan*

## 1. The preparation, characterization and Raman measurement of the graphene samples

In our experiment, we investigated free-standing graphene layers prepared by the mechanical exfoliation of kish graphite (Covalent Materials Corporation). The free-standing samples were prepared on fused quartz substrates with pre-patterned trenches (of width and depth of 4 μm) [S1]. When the suitable suspended samples were not available, we also made use of graphene samples on either bulk fused quartz or $SiO_2$(300nm)/Si substrates. All the samples were characterized by infrared spectroscopy to determine the layer thickness and stacking order [S2-S4]. We examine graphene samples with layer thickness $N$ from 1 to 6 exhibiting both Bernal (AB) and rhombohedral (ABC) stacking order.

We made use of a commercial JY Horiba Raman microscope equipped with a cooled charge-coupled device (CCD) array as the detector. We used laser excitation sources with photon energies over a wide spectral range from visible to near-infrared, including wavelengths (energies) of 488 nm (2.54 eV), 532 nm (2.33 eV), 594 nm (2.09 eV), 633 nm (1.96 eV), and 785 nm (1.58 eV). In order to record weak Raman features, we collected data for over one hour, while maintaining a low power for the excitation laser. In particular, the laser power was consistently maintained below 3 mW, with a spot size of a few microns on the sample. In this regime, heating effects were not significant. For all types of samples, the measured Raman shifts changed by less than 2 $cm^{-1}$ when laser powers were varied over the range of 0.1 – 3 mW.

## 2. Absence of LOZA, LOZO' and 2ZO Raman modes in monolayer graphene

As shown in Fig. S1, we observed two Raman peaks in the spectral range 1800 - 2050 $cm^{-1}$ for monolayer graphene (1LG). These peaks have been investigated by previous researchers and assigned as the TA-LO and TO-LA combination modes [S5, S6]. One might argue that the three additional features identified in this work as the LOZA, LOZO' and 2ZO modes in few-layer graphene (FLG) could be shifted in frequency in 1LG and then hidden within the TA-LO and TO-LA features. To exclude this possibility, we examined the spectra of free-standing 1LG for different excitation photon energies ($E_{exc}$) (Fig. S1). As the laser energy increases from 1.58 eV (785 nm) to 2.33 eV (532 nm), the two peaks blue-shift from 1755 to 1853 $cm^{-1}$ and from 1827 to 1976 $cm^{-1}$, which corresponds to a dispersion of 130 and 200 $cm^{-1}$/eV, respectively. This strongly dispersive behavior reflects that strong dispersion present in the TA and LA acoustic branches. With the TA-LO and TO-LA modes shifted away from their original positions, we did not observe any other peaks in 1800 - 2050 $cm^{-1}$ range. We can therefore safely exclude the existence of LOZA, LOZO' and 2ZO Raman modes in free-standing 1LG. Furthermore, we did not observe significant differences between the spectra of suspended 1LG and 1LG deposited on quartz substrates (Fig. S1). This indicates that the absence of LOZA, LOZO' and 2ZO Raman modes in 1LG remains robust even in the presence of a substrate.

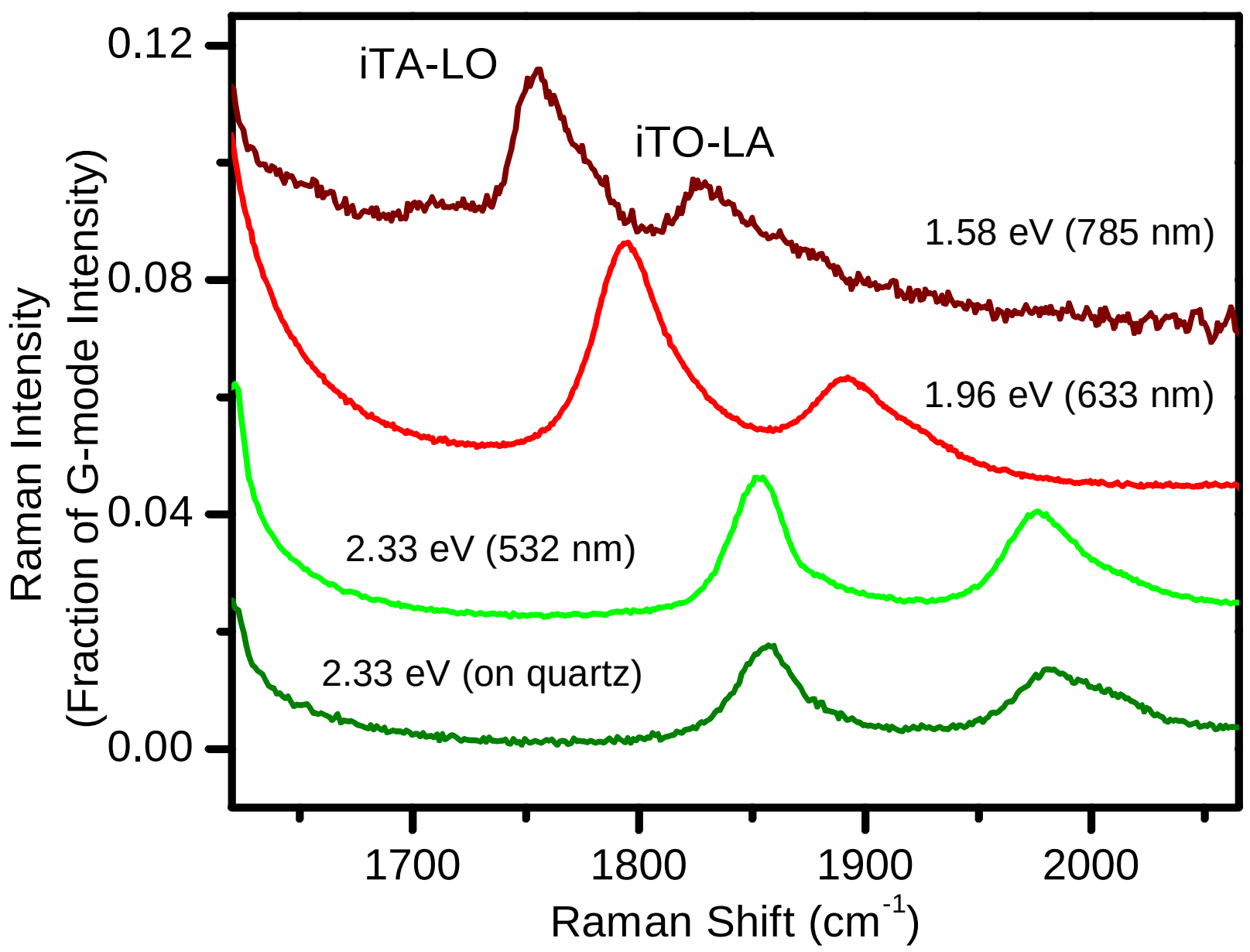

Figure S1. Raman modes in the range of 1625 to 2050 $cm^{-1}$ for free-standing monolayer graphene at laser excitation energies $E_{exc}$ (wavelengths) of 2.33 eV (532 nm), 1.96 eV (633 nm) and 1.58 eV (785 nm). The spectra are normalized with respect to the G-mode peak intensity. To evaluate the substrate effect, the spectrum of a monolayer graphene sample supported on quartz surface for $E_{exc}$ = 2.33 eV is also shown. As the substrate broadens the G-mode feature, this spectrum is normalized so that its G-mode spectrum has the same integrated intensity as that of the free-standing graphene. For clarity, the spectra are displaced successively by 0.04 units.

## 3. Measurement of 2ZO' and 2LO mode in bilayer graphene

As part of our analysis of the combination modes, we make comparison with independent data for the frequencies of the ZO' and LO modes in graphene samples. Here we briefly present experimental results for these modes that are used for our discussion. The ZO' mode is not accessible by first-order

Raman scattering [S7], so we search for the overtone or 2ZO' mode. We also examine the 2LO overtone mode derived from an intra-valley double-resonance scattering process (also designated as the 2D' mode). The use of the overtone Raman response for this mode allows us to experimentally determine the phonon frequency away from the zone center, as is appropriate for the analysis of the combination modes under consideration.

The 2ZO' mode was measured using free-standing bilayer graphene (2LG) samples in an argon-purged environment to avoid the Raman background of the substrate and the rotational Raman lines of the air molecules for Raman shifts below 200 $cm^{-1}$. We found that 2ZO' mode exhibits a measurable dependence $E_{exc}$ arising from the double-resonance mechanism [Fig. S2(a)]. In particular, we observed two components (2ZO'$^{+}$ and 2ZO'$^{-}$) in the 2ZO' line. We attribute these features, respectively, to intra-band electronic resonances within the high-lying and low-lying conduction (or valence) band in the same valley in bilayer graphene. The extracted ZO' phonon energies as a function of phonon momentum match well with theoretical predictions [S8, S9] [open triangles in Fig. 3(a) of the main text].

The spectra of the 2LO mode were measured near 3240 $cm^{-1}$ for supported bilayer graphene under ambient conditions [Fig. S2(b)]. The 2LO mode also exhibits appreciable dispersion associated with the double-resonance excitation mechanism. In our analysis, we consider the relevant LO phonon momentum probed in the measurements to be given as the average of the dominant momenta in the intra-valley resonance processes for the bilayer. The inferred LO phonon energies as a function of phonon momentum also agree well with theoretical predictions [S8, S9] [blue squares in Fig. 3(a) of the main text].

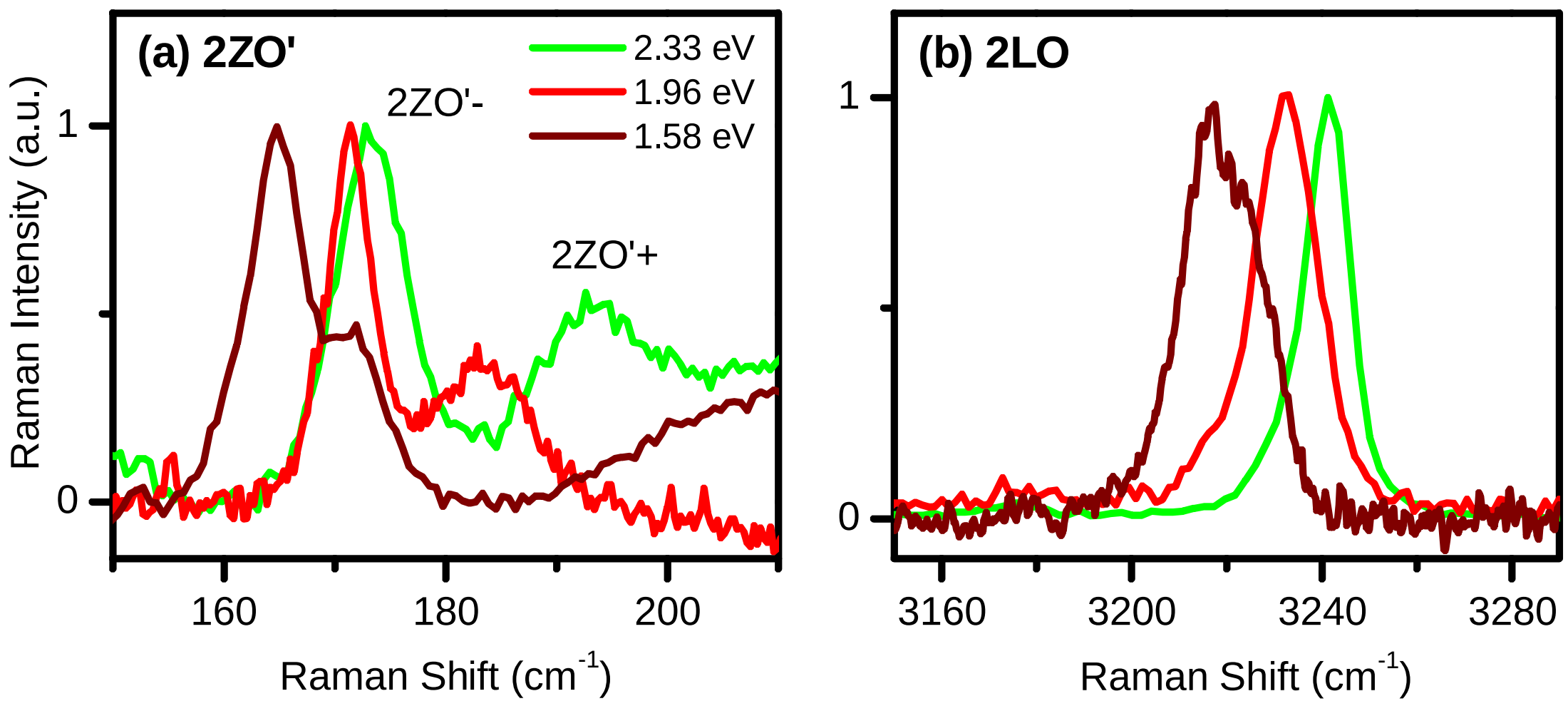

Figure S2. Raman spectra of 2ZO' mode (a) and of the 2LO mode (b) for bilayer graphene at the indicated laser excitation energies $E_{exc}$. The two peaks in the 2ZO' spectra are denoted as 2ZO'+ and 2ZO'-.

## 4. Assignment of LOZA and 2ZO mode in bilayer graphene

Here we consider in more detail the assignment of peaks P1 and P3 observed near the position of P2, which arises, as argued in this paper, from the LOZO' combination mode. Let us first consider P3. As can be seen in Fig. 2 of the main text, P3 shows weak dispersion as a function of the excitation photon energy $E_{exc}$. The frequency of P3 remains largely unchanged for excitation energies below 2.09 eV, and red shifts only slightly for higher energies. We assign P3 as the overtone of ZO (or oTO) phonon (~885 $cm^{-1}$) because both its frequency and energy dependence match the ZO phonon branch, which is flat near Γ-point and shows a slight negative dispersion with increasing phonon momentum from the Γ-point. The same assignment of 2ZO has been made on the corresponding Raman mode in carbon nanotubes (the so-called M-mode) [S10].

Fig. 2 of the main text also displays the dispersion of P1, which is found to blue-shift with increasing excitation photon energy, with average dispersion of 39 $cm^{-1}$/eV. We assign P1 to the LO+ZA combination mode for phonons near Γ-point, as accessed by the intra-valley double-resonance process. Since the ZA phonons cannot be accessed directly in our Raman experiment due to their low energy, we made use of the theoretical values for ZA-phonon energy [S8, S9] [open diamonds in Fig. 2(b) based on the ZA branch in Fig. 3(a)]. As shown in Fig. 2(b), the LO+ZA combination (open green circles) match well with the frequency and dispersion of P3 (solid green dots). On the other way, we extract the ZA phonon energies by subtracting the experimental LO-mode energies from the experimental LOZA-mode energies. The obtained ZA phonon energies compare well with the theoretical ZA dispersion [Fig. 3(a)]. These arguments support our assignment of P1 as the LOZA combination mode.

## 5. Behavior of LOZO' mode for graphene trilayers with ABA and ABC stacking order

We consider here in more detail the dependence of the LOZO' mode on crystallographic stacking in trilayer graphene. In Fig. S3(a), we show the LOZA, LOZO' and 2ZO Raman peaks in trilayer graphene with both ABA (Bernal) and ABC (rhombohedral) stacking order for different excitation laser energies. For all laser photon energies, the LOZO' Raman band appears as a single peak for ABA trilayers, but is split into three well-separated narrow peaks for ABC trilayers. The frequencies of these sub-peaks as a function of $E_{exc}$ are plotted in Fig. S3(b), along with the dispersions of the LOZA and 2ZO modes. In the case of ABC trilayers, the frequencies of all three sub-peaks of the LOZO' mode blue-shift with the increasing $E_{exc}$. The dispersions of all three sub-peaks are similar to one another and also to the case of bilayer graphene.

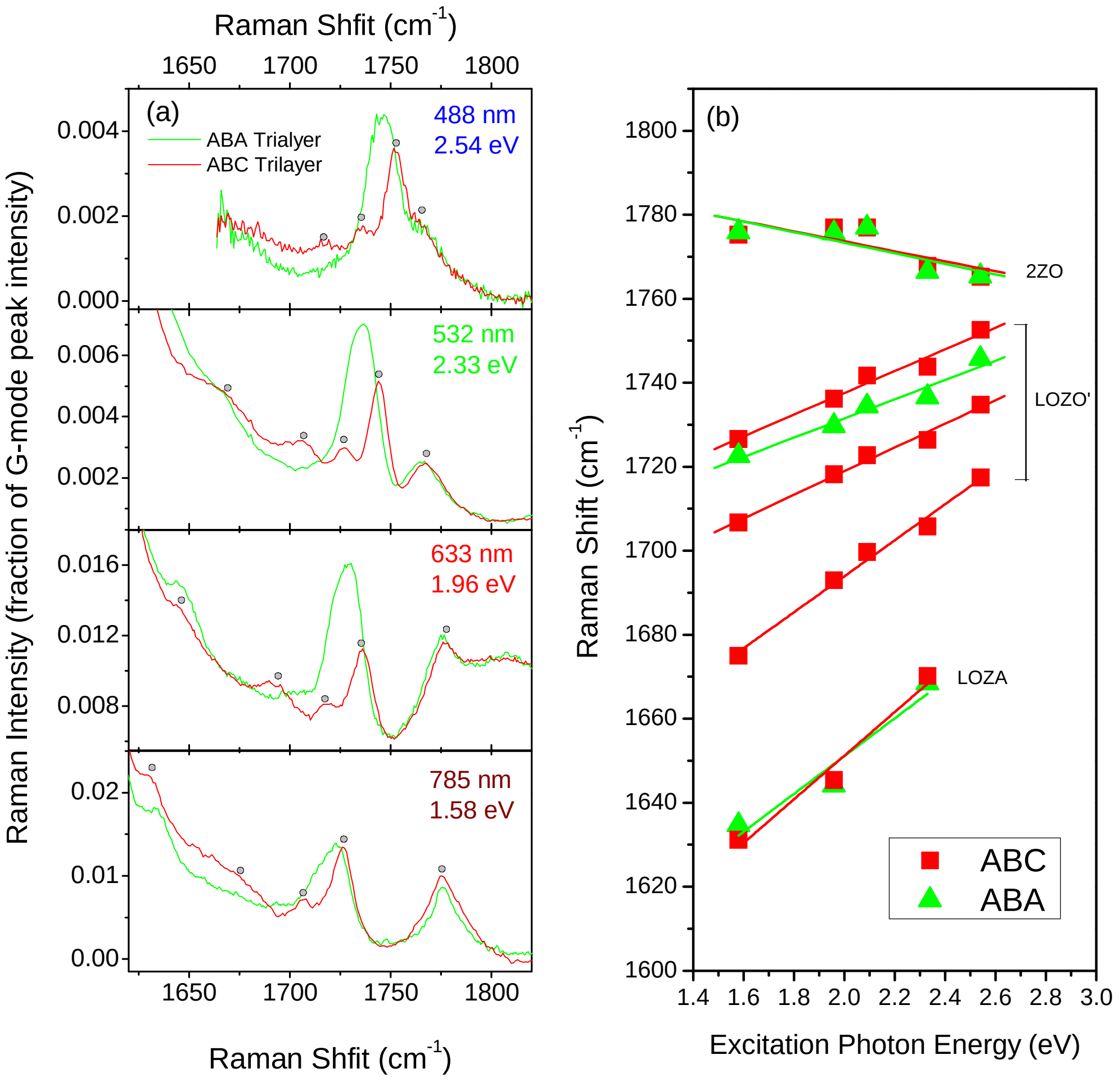

Figure S3. (a) Raman spectra for trilayer graphene with ABA (green) and ABC (red) stacking order taken over the range of 1620 to 1820 $cm^{-1}$ for the indicated laser excitation energies $E_{exc}$. (b) Raman dispersion data based on the peak positions indicated in (a). The green triangles and the red rectangles are for ABA and ABC trilayers, respectively. The lines are the linear fits of the data.

## Supplemental References